\documentclass[twocolumn,aps,showpacs,superscriptaddress,floatfix]
{revtex4}
\usepackage[dvips]{graphicx}
\newcommand{\tr}{{\rm \, Tr }\, }

\newcommand{\beq}{\begin{equation}}
\newcommand{\eeq}{\end{equation}}
\newcommand{\beqa}{\begin{eqnarray}}
\newcommand{\eeqa}{\end{eqnarray}}
\newcommand{\beqan}{\begin{eqnarray*}}
\newcommand{\eeqan}{\end{eqnarray*}}

\begin{document}
\title{Controllable generation of highly nonclassical states 
from nearly pure squeezed vacua}

\author{Kentaro Wakui}
\email{kwakui@nict.go.jp}
\affiliation{
    National Institute of Information and Communications 
    Technology, 
    4-2-1 Nukui-Kita, Koganei, Tokyo 184-8795, Japan}
\affiliation{
    Department of Applied Physics, The University of Tokyo 
    7-3-1 Hongo, Bunkyo-ku, Tokyo 113-8656, Japan}
\affiliation{
    CREST, Japan Science and Technology Agency, 
    1-9-9 Yaesu, Chuoh, Tokyo 103-0028, Japan}

\author{Hiroki Takahashi}
\affiliation{
    National Institute of Information and Communications 
    Technology, 
    4-2-1 Nukui-Kita, Koganei, Tokyo 184-8795, Japan}
\affiliation{
    Department of Applied Physics, The University of Tokyo 
    7-3-1 Hongo, Bunkyo-ku, Tokyo 113-8656, Japan}
\affiliation{
    CREST, Japan Science and Technology Agency, 
    1-9-9 Yaesu, Chuoh, Tokyo 103-0028, Japan}

\author{Akira Furusawa}
\affiliation{
    Department of Applied Physics, The University of Tokyo 
    7-3-1 Hongo, Bunkyo-ku, Tokyo 113-8656, Japan}
\affiliation{
    CREST, Japan Science and Technology Agency, 
    1-9-9 Yaesu, Chuoh, Tokyo 103-0028, Japan}

\author{Masahide Sasaki}
\email{psasaki@nict.go.jp}
\affiliation{
    National Institute of Information and Communications 
    Technology, 
    4-2-1 Nukui-Kita, Koganei, Tokyo 184-8795, Japan}
\affiliation{
    CREST, Japan Science and Technology Agency, 
    1-9-9 Yaesu, Chuoh, Tokyo 103-0028, Japan}
%
%
%\date{\today} 
\begin{abstract}
We present controllable generation of 
various kinds of highly nonclassical states of light, 
including the single photon state and 
superposition states of mesoscopically distinct components. 
The high nonclassicality of the generated states is 
measured by the negativity of the Wigner function, 
which is largest ever observed to our knowledge. 
Our scheme is based on photon subtraction 
from a nearly pure squeezed vacuum, 
generated from an optical parametric oscillator with 
a periodically-poled KTiOPO$_4$ crystal as a nonlinear medium. 
This is an important step to realize basic elements of 
universal quantum gates, 
and to serve as a highly nonclassical input probe 
for spectroscopy and the study of quantum memory. 
\end{abstract}
%
% 42.50.Dv Nonclassical states of the electromagnetic field, 
%          including entangled photon states; 
%          quantum state engineering and measurements 
% 03.65.Wj State reconstruction, quantum tomography 
% 03.67.Mn Quantum entanglement production, characterization, 
%          and manipulation 
% 03.67.Pp Quantum error correction and other methods for 
%          protection against decoherence 
% 42.50.Md Optical transient phenomena: 
%          quantum beats, photon echo, free-induction decay, 
%          dephasings and revivals, optical nutation, 
%          and self-induced transparency 
% 78.20.Bh Optical properties, condensed matter spectroscopy 
%          and other interaction of radiation and particles with 
%          condensed matter: 
%          Theory, models, and numerical simulation
%
%\date{\today}
\maketitle

%%%%%%%%%%%%%%%%%%%%%%%%%%%%%%%%%%%%%%%%%%%%%%%%%%%%%%%%%%%
\section{Introduction}
\label{Intro}
%%%%%%%%%%%%%%%%%%%%%%%%%%%%%%%%%%%%%%%%%%%%%%%%%%%%%%%%%%%

%{\bf Introduction}
Controllable generation of various kinds of nonclassical states 
of light is crucial for the study of fundamental aspects of 
quantum mechanics and developing quantum information science. 
One of effective methods is a conditional preparation scheme 
by photon counting on one part of an entangled state 
produced in the parametric down-conversion (PDC)\,%
\cite{Yurke_Stoler87,Dakna97}. 
The main feature of the quantum states thus prepared is that 
their Wigner functions $W(x, p)$, 
which is a quasi-probability distribution 
for non-commuting quadrature observables 
$\hat x \hat p - \hat p \hat x = i$, 
should exhibit the negative values in certain phase-space regions. 
This is in sharp contrast to that 
a squeezed state, 
which are another representative of nonclassical states, 
have a non-negative definite Gaussian distribution 
of the Wigner function.

Generation and observation of nonclassical states 
with negative values of $W(x, p)$ 
have been reported recently. 
Those states can be categorized into two families. 
One family is the number states and their variants 
combined with coherent state components\,% 
\cite{Single_photon_Lvovsky_Zavatta,Ourjoumtsev_etal_qph06,Variants_single_photon_Lvovsky_Zavatta}, 
created in a non-colinear PDC configuration 
of the signal and idler photons. 
The other is the photon subtracted squeezed states, 
where a small fraction of a squeezed vacuum beam 
is beam-split and guided into a photon counter 
as trigger photons, 
and the remaining signal beam is output 
conditioned by the detection of the trigger photons\,%
\cite{Dakna97}. 
These states must ideally be a particular superposition 
of odd number of photons, 
and can constitute superposition states of mesoscopically 
distinct components, sometimes referred to as 
the Schr\"{o}dinger kittens\,%
\cite{Ourjoumtsev_etal_Science06_Kitten,Jonas_PRL06_Kitten}.

The negativity of the Wigner function can be easily lost 
with experimental imperfections. 
Among them, the imperfections of the homodyne detector 
are often corrected to reconstruct a Wigner function 
for a generated state just in front of the homodyne detector. 
Direct measurement of the negativity of the Wigner function 
requires a generator with low loss and low noise 
as well as a nearly ideal homodyne detector. 
Negative values for the uncorrected Wigner functions were 
marked, e.g. 
$W(0, 0)=-0.026$ for the Schr\"{o}dinger kittens 
in the photon subtraction scheme\,%
\cite{Ourjoumtsev_etal_Science06_Kitten}, 
and 
$W(0, 0)=-0.04$ for the single photon state 
in the non-colinear PDC scheme. 
(The values correspond to a unit convention according to 
$\hat x \hat p - \hat p \hat x = i$.)

We present a scheme that can generate 
the photon subtracted squeezed states 
with the deepest negative dips of the Wigner functions 
ever observed to our knowledge, 
including the single photon state and the Schr\"{o}dinger kittens. 
Realization of the single photon regime 
particularly requires a nearly pure squeezed input 
so as to collect appropriate trigger photons 
with high signal-to-noise ratio 
from a small fraction (10\% at most) of the squeezed beam, 
which failed in the previous photon subtraction schemes. 
Our scheme enables one to vary the regime from 
the single photon to the Schr\"{o}dinger kittens 
by simply tuning the squeezing level 
which is directly controlled by a pump power for the squeezer, 
and hence to study continuous transition 
from the single photon regime 
to the mesoscopic superposition regime.

%%%%%%%%%%%%%%%%%%%%%%%%%%%%%%%%%%%%%%%%%%%%%%%%%%%%%%%%%%%
\section{Experiment}
\label{Experiment}
%%%%%%%%%%%%%%%%%%%%%%%%%%%%%%%%%%%%%%%%%%%%%%%%%%%%%%%%%%%

%%%%%%%%%%%%%%%%%%%%%%%%%%%%%%%%%%%%%
\begin{figure}
\hspace{10mm}
\begin{center}
\includegraphics[width=0.95\linewidth]{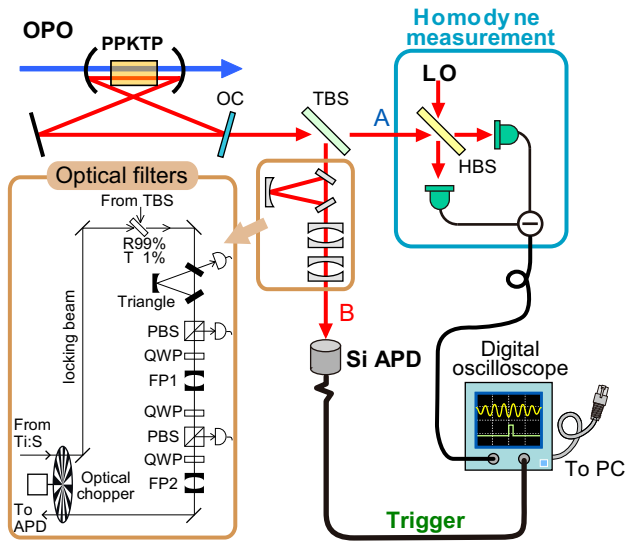}
\caption{Schematic of experimental setup. 
OC:output coupler, TBS:tapping beam splitter, 
Triangle:triangle cavity, PBS:polarizing beam splitter, 
QWP:quarter wave plate, FP:Fabry-Perot cavity, 
HBS:50:50 beam-splitter.}
\label{fig:Exp_setup}
\end{center}
\end{figure}
%%%%%%%%%%%%%%%%%%%%%%%%%%%%%%%%%%%%%

%{\bf Experiment} 
The key is an optical parametric oscillator (OPO) 
with a periodically-poled KTiOPO$_4$ (PPKTP) crystal 
as a nonlinear medium, 
which can generate nearly pure squeezed vacuum states 
in a continuous-wave (cw) regime\,%
\cite{Aoki_etal_OE06,Suzuki_etal_APL06}. 
A schematic of our experimental setup is shown in 
Fig.\,\ref{fig:Exp_setup}, 
which is similar to the previous work\,%
\cite{Jonas_PRL06_Kitten}. 
A continuous-wave Ti:Sapphire laser (Coherent MBR110) is used 
as a primary source of the fundamental beam at 860\,nm,  
which is mainly used to generate 
second harmonic (430\,nm) of about 200\,mW 
by a frequency doubler (a bow-tie cavity with KNbO$_3$), 
and is also used  as a local oscillator (LO) 
for homodyne detection, 
and probe beams for various control purposes. 
The second harmonic beam is used to pump 
the OPO with a 10\,mm long PPKTP crystal (Raicol Crystals) 
in an optical cavity (a bow-tie configuration 
with a round-trip length of about 520\,mm). 
An output coupler (OC) of this squeezer cavity 
has a transmittance of 10.3\,\%, 
and the intracavity loss is about 0.2$\sim$0.3\,\%. 
The FWHM of the cavity is about 9.3\,MHz.

\begin{figure}
\hspace{10mm}
\begin{center}
\includegraphics[width=0.8\linewidth]{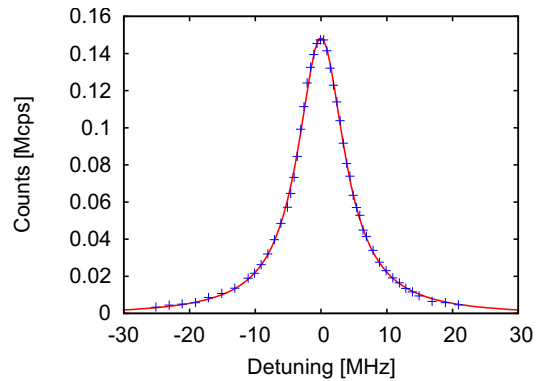}
\caption{Photon counting spectrum of the trigger photons 
through the optical filtering cavities 
around the degenerate frequency as a function of cavity detuning: 
solid crosses for experiment, solid line for theory. 
The pump power was 160mW. All the OPO output is fed into the APD path
for this experiment.} 
\label{fig:SqzDetuned}
\end{center}
\end{figure}

A small fraction (1$\sim$10\,\%) of the squeezed vacuum beam 
in path A is tapped at a beam splitter (TBS), 
guided into a commercial Si-APD (Perkin Elmer SPCM-AQR-16) 
through three optical filtering cavities in a row in path B, 
and is used as trigger signals for conditional photon subtraction. 
Figure\,\ref{fig:SqzDetuned} shows a photon counting spectrum 
of the trigger photons through the filtering cavities, 
which have 5$\sim$10 times wider bandwidths than that of the OPO, 
measured by detuning the OPO 
around the fundamental, degenerate, wavelength 860\,nm. 
This consists of a single peak with a bandwidth of 8.6\,MHz (FWHM), 
which can be easily matched with the observable mode 
by the homodyne detector. 
Other irrelevant, nondegenerate, modes from the OPO, 
peaking at every free spectral range of 573\,MHz 
apart from the degenerate frequency, 
are well suppressed sufficiently. 
%with an extinction ratio about 120\,dB. 
The total transmission ratio for the mode of interest 
is about 30\,\% just in front of the Si-APD. 
The trigger counting rate varies from less than 1\,kcps up to 50\,kcps, 
changing with the pump power of OPO up to 160\,mW and tapping ratios,
These trigger counting rates are mostly much greater than 
the Si-APD dark counts (100cps). 
%%%%%%%%%%%%%%%%%%%%%%%%%%%%%%%%%%%%%%
%%%%%%%%%%%%%%%%%%%%%%%%%%%%%%%%%%%%%%

The filter cavities are locked 
by a ``sample-and-hold locking" technique 
which enables us to switch the system 
from a ``locking phase" to a ``measurement phase" periodically. 
In the locking phase, 
the filter cavities are locked in a conventional manner 
(FM-sideband locking technique). 
On the other hand in the measurement phase, 
the locking beam is cut off for photon counting 
and 
servo amplifiers hold the system in the same state 
as the one right before the locking beam is cut off. 
We attain this periodically to have the locking beam 
of the filters pass one optical chopper two times,
before injected into filter cavities and after transmitted them before APD.
Shuttered timings are shifted each other in half period 
with respect to chopping frequency\,(500Hz). 
Uneven duty cycles of chopping disk prevent the locking beam 
from entering the Si-APD all the time. 
We also rebuild our servo amplifiers in order to be able 
to accept external timing signal
and switch the two phases synchronized to the optical chopper's
driver.

The generated nonclassical light is combined with the LO 
at a 50:50 beam-splitter (HBS) 
and detected by a balanced homodyne detector 
with Si photodiodes 
(Hamamatsu S-3759, anti-reflection coated at 860\,nm, 
99.6\,\% quantum efficiency). 
In order to improve the homodyne efficiency, 
the LO beam is spatially filtered by the mode cleaning cavity 
which yields the same spatial mode as the OPO output. 
The propagation loss (1$\sim$10\,\%) mainly 
comes from the tapping itself, 
and the homodyne efficiency is 98\,\%.

%%%%%%%%%%%%%%%%%%%%%%%%%%%%%%%%%%%%%%%
\begin{figure*}[t]
\center{\includegraphics[width=0.98\linewidth]
{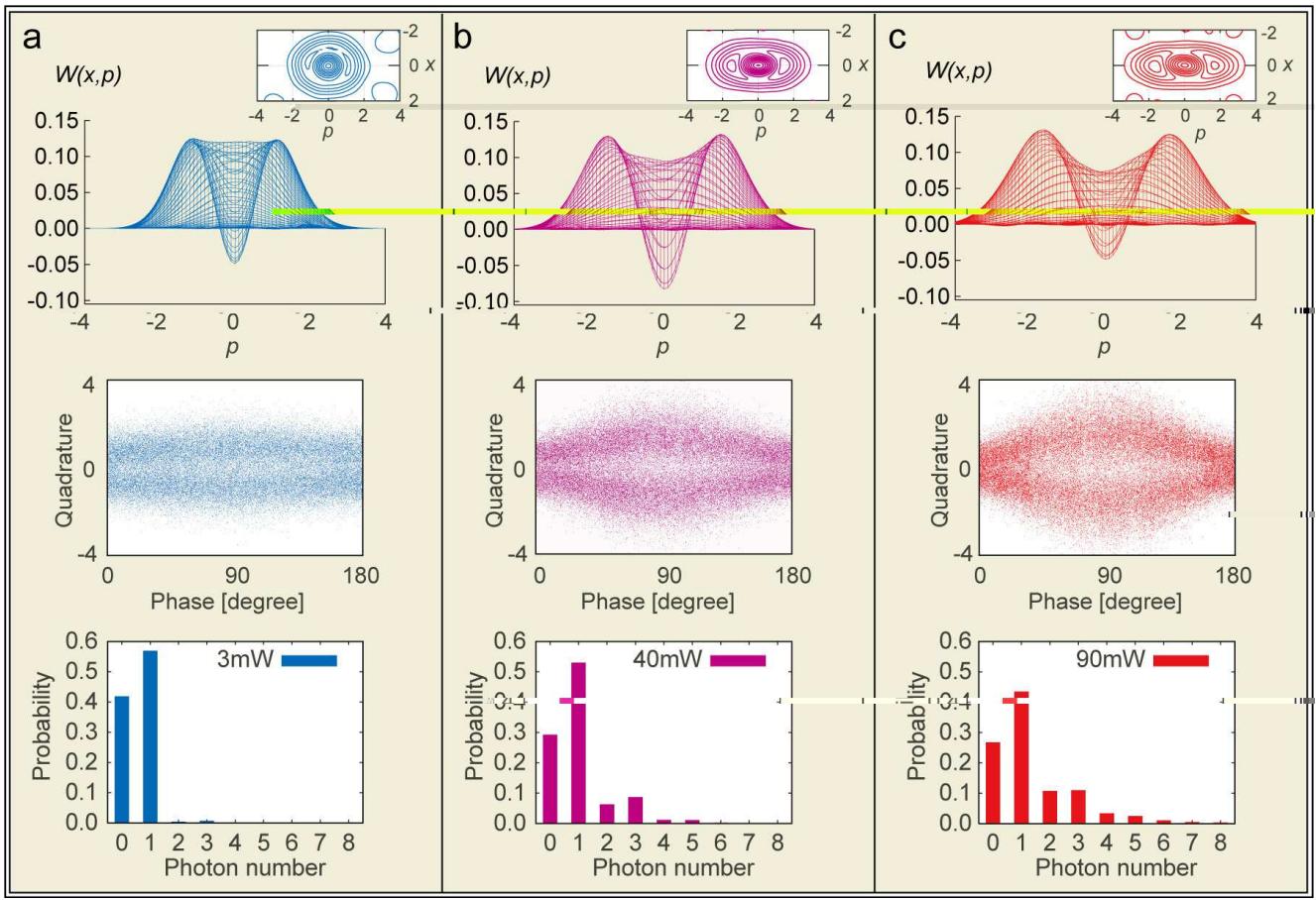}}
\caption{
Experimental Wigner functions (top panels) 
constructed from raw data 
without any correction of measurement imperfections 
in the case of 5\% tapping ratio. 
(a) Single photon state generated by -0.7dB squeezed input. 
(b) and (c) Schr\"{o}dinger kittens generated 
by -2.6dB and -3.7dB squeezed inputs, respectively. 
The insets in top panels are the contours of the Wigner functions. 
The middle panels are quadrature distributions obtained 
by homodyne detection. 
The bottom panels are photon number distributions 
obtained by the iterative maximum-likelihood estimation. } 
\label{fig:Wigner_functions_wave_forms}
\end{figure*} 
%%%%%%%%%%%%%%%%%%%%%%%%%%%%%%%%%%%%%%%

For every trigger signal from the Si-APD, 
a digital oscilloscope (LeCroy WaveRunner 6050A) 
samples homodyne signals over a period $\sim$0.5\,$\mu$s 
around the time when the trigger signal is detected, 
and sends them to a PC one after another. 
Each segment of the homodyne signals are then 
time-integrated, being multiplied 
by a particular temporal mode function $\Psi_0(t)$, 
to provide the quadrature amplitude finally observed 
with respect to which the Wigner functions are constructed. 
About 50,000 data points of the quadrature amplitude are collected, 
and the Wigner function is constructed 
using the iterative maximum-likelihood estimation algorithm\,%
\cite{Lvovsky_JOB04_MaxLike}.

%%%%%%%%%%%%%%%%%%%%%%%%%%%%%%%%%%%%%%%%%%%%%%%%%%%%%%%%%%%
\section{Result and analysis}
\label{Result_and_analysis}
%%%%%%%%%%%%%%%%%%%%%%%%%%%%%%%%%%%%%%%%%%%%%%%%%%%%%%%%%%%

%{\bf Result and analysis} 
The temporal mode function $\Psi_0(t)$ should be chosen 
such that it defines the signal mode 
which shares the maximal entanglement 
with the trigger photon mode. 
In the Si-APD, 
a projection onto photon number states takes place 
in a time scale $\alt400$\,ps. 
The trigger outputs in this time scale 
are sent into the digital oscilloscope 
to sample homodyne signals. 
The time resolution in this sampling is about $T=1$\,ns. 
This is slightly longer than the Si-APD time scale, 
but much shorter than 
the temporal correlation of the squeezed vacua 
whose bandwidth is typically $2B\sim10$\,MHz. 
This $T$ finally defines the trigger photon mode, 
which can simply be a rectangular temporal mode function. 
For such a small $BT$, 
a single mode description is valid\,%
\cite{Sasaki_Suzuki_PRA06}.  
In a good approximation, 
one can consider $\Psi_0(t)$ 
in a form\,%
\cite{Molmer_PRA06} 
(see also Appendix \ref{Appendix:Mode functions}) 
%
%$\Psi_0(t) =\sqrt{\zeta_0}e^{-\zeta_0\vert t\vert}$, 
%  
\beq\label{Psi_0}
\Psi_0(t) =\sqrt{\zeta_0}e^{-\zeta_0\vert t\vert},  
\eeq
assuming a trigger signal detected at $t=0$, 
where $\zeta_0\equiv(\gamma_T+\gamma_L)/2$ 
determines the characteristic bandwidth 
$\zeta_0/\pi\sim9.3$\,MHz of the cavity 
for the leakage rates  
$\gamma_T\sim57$\,MHz of the output coupler 
and 
$\gamma_L\sim1.2$\,MHz of the cavity loss.

Figure\,\ref{fig:Wigner_functions_wave_forms} shows 
experimental Wigner functions (top panels), 
quadrature distributions over half a period (middle panels), 
and photon number distributions (bottom panels) 
of the photon subtracted squeezed states. 
None of correction of measurement imperfections is made. 
From the left, 
the single photon state (Fig.\,\ref{fig:Wigner_functions_wave_forms}a) 
by -0.7dB squeezed input, 
and Schr\"{o}dinger kittens with two kinds of amplitudes 
(Fig.\,\ref{fig:Wigner_functions_wave_forms}b 
by -2.6dB squeezed inputs
and\,\ref{fig:Wigner_functions_wave_forms}c 
by -3.7dB squeezed input). 
The tapping fraction of TBS in 
Fig. \ref{fig:Exp_setup} is set to 5\,\%. 
The large negativity is obtained 
in a wide range of squeezing levels.

%%%%%%%%%%%%%%%%%%%%%%%%%%%%%%%%%%%%%%
\begin{figure}
\hspace{10mm}
\begin{center}
\includegraphics[width=1.0\linewidth]{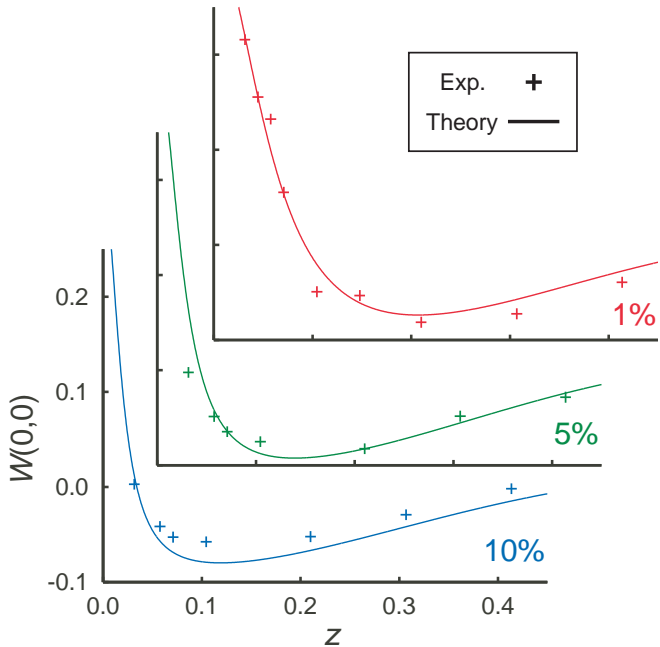}
\caption{The dependence of $W(0,0)$ on the squeezing level.}
\label{fig:Origin_Wigner_functions}
\end{center}
\end{figure}
%%%%%%%%%%%%%%%%%%%%%%%%%%%%%%%%%%%%%%

Figure\,\ref{fig:Origin_Wigner_functions} shows 
the values of the Wigner function at the phase-space origin 
$W(0,0)$ (solid circles) 
as a function of the ratio of the pump light amplitude 
to the threshold amplitude of the OPO, 
$z=\sqrt{P_{pump}/P_{th}}$, 
which directly connected to the squeezing level 
in the cases of three tapping ratios 1\%, 5\%, and 10\%. 
The lines are fittings by a theoretical model 
based on the two-mode state reduced to 
the signal mode $\Psi_0(t)$ 
and the trigger photon mode $\phi_0(t)$ 
(Appendix \ref{Appendix:Formula of Wigner function}).

The model includes two kinds of unwanted losses. 
One is in the homodyne channel modeled by 
a transmittance $\tau_h=$78\,\%, 
taking imperfections of homodyne electronics into account. 
The other is a squeezing-level-dependent loss, 
$\tau_s(z)=\tau_{s0}-\kappa z^2$, 
which also include a constant linear loss 
of passive optical elements. 
The latter is necessary to explain the degradation of 
the negativity of $W(0,0)$ as $z$ increases. 
A loss induces a mixing of even-number of states into 
the photon subtracted squeezed state 
which ideally contains only odd-number states. 
Since the values of $W(0,0)$ for even-number of states 
are positive, 
the negativity of $W(0,0)$ is easily destroyed by a loss. 
The $z$-dependence of $W(0,0)$ cannot be explained by 
any model with constant linear loss.  
The optimal fitting is attained for 
$\tau_{s0}=0.95$ and $\kappa=0.93$.

One of main causes for $z$-dependent loss might be 
phase fluctuations of the LO in the homodyne detector 
and of the locking beam at several cavities. 
Such fluctuations cause effectively phase noises 
which becomes larger for larger radius in the phase space. 
As $z$ increases, 
the Wigner function spreads over the phase space, 
and suffers from such fluctuations. 
For realizing larger Schr\"{o}dinger kittens, 
it will also be important to suppress phase fluctuations.

In the single photon regime, 
i.e., small $z$, 
the trigger count rate becomes small. 
A primary factor to determine the behavior of $W(0,0)$ 
is then the noise count rate $\nu$ in the trigger channel, 
which include the dark counts of the Si-APD and background photons.  
It is necessary to reduce this rate to generate 
the single photon state with a deep negative Wigner function. 
In this regard, 
the OPO with PPKTP works very well to reduce noisy photons.

%%%%%%%%%%%%%%%%%%%%%%%%%%%%%%%%%%%%%%%%%%%%%%%%%%%%%%%%%%%
\section{Concluding remarks}
\label{Concluding_remarks}
%%%%%%%%%%%%%%%%%%%%%%%%%%%%%%%%%%%%%%%%%%%%%%%%%%%%%%%%%%%

%{\bf Concluding remarks} 
We have presented 
the photon subtracted squeezed states 
with large negative dips in their Wigner functions. 
Those highly nonclassical states 
cover the single photon regime and 
the Schr\"{o}dinger kitten regime.  
Their continuous transition can be 
controlled by the squeezing level of the input source. 
The nearly pure input source from the OPO with PPKTP 
enabled us to study the degradation of the negative 
Wigner functions in detail. 
It was found that there was 
the squeezing-level-dependent loss, 
whose origins remains open, 
but we suspect at present 
that it stems from phase fluctuations 
of the LO and the locking beam.

Our scheme will particularly be useful 
to make the output states 
coupled to atomic media for quantum memory, 
and also to apply them to spectroscopic applications. 
In addition, our scheme can directly be combined 
with well-matured technology of Gaussian operations 
on quadrature observables in a continuous-wave scheme\,%
\cite{Yonezawa_etal_Nature04,Takei_etal_PRL05} 
for pursuing further challenges. 
For example, the attained level of the nonclassicality, 
i.e. the large negativity, 
will allow one to teleport the nonclassicality 
of non-Gaussian input states, 
which remains an unreached target in spite of its importance 
in quantum information science, 
by using the presently available squeezer with -7dB squeezing\,%
\cite{Suzuki_etal_APL06,Braunstein_KimblePRL98_Teleport}.

Another important challenge is to combine 
photon-number resolving detectors\,%
\cite{Waks_etal_PRL04_VLPC4,Rosenberg_etal_PRA05_Supercond_TES_NIST2,Fujiwara_Sasaki_OL06} 
with our scheme 
for breeding the Schr\"{o}dinger kittens into cats, 
and for developing a universal quantum gate circuit  
both for discrete and continuous variables\,%
\cite{Gottesman01,BartlettSanders02,Sasaki_etal_QCMC04}.

\begin{acknowledgements}
 Authors thank K. Hayasaka, J. S. Neergaard-Nielsen, T. Aoki, 
 H. Yonezawa, S. Suzuki and M. Takeoka for their valuable 
comments and suggestions.  
\end{acknowledgements}

\appendix

%%%%%%%%%%%%%%%%%%%%%%%%%%%%%%%%%%%%%%%%%%%%%%%%%%%%%%%%%%%%%%%
\section{Mode functions}
\label{Appendix:Mode functions}
%%%%%%%%%%%%%%%%%%%%%%%%%%%%%%%%%%%%%%%%%%%%%%%%%%%%%%%%%%%%%%%

To describe the trigger photon mode localized in the time domain, 
an appropriate basis is given by 
\beq\label{DFT functions}
\phi_k(t)
=
\left\{
\begin{array}{ll}
\displaystyle\frac{1}{\sqrt{T}} 
\mathrm{exp}(-i\displaystyle\frac{2\pi kt}{T}) 
& \quad 
 -\displaystyle\frac{T}{2}
  \le t \le 
  \displaystyle\frac{T}{2}, 
\\
0 
& \quad 
\mathrm{otherwise},
\end{array}\right. 
\eeq
In the present case where $BT\ll1$, 
the lowest mode 
is excited in the Si-APD with the weight $\sim BT$, 
and the weights of the other higher modes are negligible\,%
\cite{Sasaki_Suzuki_PRA06}.  
The trigger mode function is thus $\phi_0(t)$. 
The total quantum efficiency for this mode is given by 
$\eta=\eta_0\eta_F BT$ 
where $\eta_0$ is the intrinsic quantum efficiency 
and $\eta_F$ is the transmittance through the optical filters. 
Taking a noise count rate $\nu$ further into account, 
detection of the trigger photons is described 
by the on/off detector POVM as specified in ref.\,%
\cite{Sasaki_Suzuki_PRA06}.

The signal mode $\Psi_0(t)$ should be appropriately determined 
by the statistics of the conditional output state 
$\hat\rho_A$ in path A given by 
\beq
\hat\rho_A=
\frac{ \tr_B[\hat\rho_{AB}
             \hat{\Pi}_\mathrm{on}^B(\eta,\nu)] }
     { \tr_{AB}[\hat\rho_{AB}
                \hat{\Pi}_\mathrm{on}^B(\eta,\nu)] }, 
\eeq
where the state $\hat\rho_{AB}$ corresponds to 
the split beams in paths A and B just after the TBS 
(Fig. \ref{fig:Exp_setup}). 
We consider a mode expansion of 
the field operator to describe the state  $\hat\rho_A$ 
\beq
\hat A(t)\sim\sum_{k=0}^{K-1} \hat A_k \Psi_k(t),   
\eeq
where 
\beq\label{Orthonormality}
\int_{-\infty}^{\infty} dt \Psi_k^\ast(t) \Psi_k(t)=\delta_{kl}. 
\eeq
The mode functions $\{\Psi_k(t)\}$ are determined 
by minimizing the square error 
\beqa
J_K&=&
\tr\Biggl[
\int_{-\infty}^{\infty} dt 
\biggl\vert \hat A(t)-\sum_{k=0}^{K-1} \hat A_k \Psi_k(t) 
\biggr\vert^2 
\Biggr]
\nonumber\\
&=&
\int_{-\infty}^{\infty} dt \tr[\hat A^\dagger(t)\hat A(t)]
-
\sum_{k=0}^{K-1} 
\tr[\hat A_k^\dagger \hat A_k]. 
\eeqa
Since the first term is constant, 
we are to maximize the second term 
under the constraint of Eq. (\ref{Orthonormality}) for each $k$. 
It is known that the optimal solution for this is 
given by the eigenfunctions of the integral equation 
\beq\label{Integral equation for Psi in text}
\chi_k \Psi_{k}(t) 
=\int_{-\infty}^{\infty} dt' h(t,t') \Psi_{k}(t'),  
\eeq
where 
\beq\label{filter h}
h(t,t') =\tr_{A}[\hat\rho_{A} \hat A^\dagger(t) \hat A(t')], 
\eeq
assuming that the trigger signal is detected at $t=0$\,%
\cite{Molmer_PRA06,Sasaki_Suzuki_PRA06}.

In a good approximation, 
one can consider only the lowest order mode $\Psi_0(t)$ 
in a form 
\beq\label{Psi_0}
\Psi_0(t) =\sqrt{\zeta_0}e^{-\zeta_0\vert t\vert}.  
\eeq
The Wigner function is constructed with respect to 
the field component defined by 
\beq
\hat A
\equiv 
\hat A_0=\int_{-\infty}^{\infty} dt \hat A(t) \Psi_0(t).  
\eeq

%%%%%%%%%%%%%%%%%%%%%%%%%%%%%%%%%%%%%%%%%%%%%%%%%%%%%%%%%%%%%%%
\section{Formula of Wigner function}
\label{Appendix:Formula of Wigner function}
%%%%%%%%%%%%%%%%%%%%%%%%%%%%%%%%%%%%%%%%%%%%%%%%%%%%%%%%%%%%%%%

The Wigner function can be obtained 
by evaluating the covariance matrices 
in terms of the signal mode $\Psi_0(t)$ 
and the trigger photon mode $\phi_0(t)$ 
as prescribed in ref. \,%
\cite{Molmer_PRA06}, 
exploiting the on/off detector model in ref.\,%
\cite{Sasaki_Suzuki_PRA06}:
\beq\label{Wigner function:final}
W(x,p)
=
\frac{ R(x,p;0,0) - R(x,p;\eta,\nu) }
{1-{\cal{N}}(\eta,\nu)}, 
\eeq
where 
%\beq\label{R:final}
%R(x,p;\eta,\nu) 
%=
%\frac
%{
%{\cal{N}}(\eta,\nu)
%\displaystyle
%\exp
%\Biggl(
%-\frac{x^2}{1-\tau_h+ \tau_h \sigma^2(z)}
%-\frac{p^2}{1-\tau_h+ \tau_h \sigma^2(-z)}
%\Biggr)
%}
%{
%\pi 
%\sqrt{\left[1-\tau_h+ \tau_h \sigma^2(z) \right]
%\left[1-\tau_h+ \tau_h \sigma^2(-z) \right]}
%}, 
%\eeq
%
%
%
\beqa\label{R:final}
\lefteqn{R(x,p;\eta,\nu) } \nonumber \\
&&=
\frac
{
{\cal{N}}(\eta,\nu)
}
{
\pi 
\sqrt{\left[1-\tau_h+ \tau_h \sigma^2(z) \right]
\left[1-\tau_h+ \tau_h \sigma^2(-z) \right]}
} \nonumber \\
&&\times \exp
\Biggl(
-\frac{x^2}{1-\tau_h+ \tau_h \sigma^2(z)}
-\frac{p^2}{1-\tau_h+ \tau_h \sigma^2(-z)}
\Biggr), \nonumber \\
\eeqa
with 
\beq\label{N}
{\cal{N}}(\eta,\nu)
=
\frac
{2 e^{-\nu}}
{\sqrt{\left\{ 2+\left[ b(z)-1 \right]\eta \right\}
       \left\{ 2+\left[ b(-z)-1 \right]\eta \right\} } } , 
\eeq
and 
\beq\label{sigma2}
\sigma^2(z)
=
a(z)
-
\frac
{c^2(z)\eta}
{2+\left[ b(z)-1 \right]\eta }, 
\eeq
where 
\beq\label{a}
a(z)
=
1-\tau+\tau 
\left[1- 
\frac{2 \tau_s(z)\gamma_T}{\zeta_0}
%\cdot
\frac{z(z+3)}{(z+1)(z+2)^2}  \right], 
\eeq
\beq\label{b}
b(z)
=
\tau
+
(1-\tau)
\left(
  1-\tau_s(z) \gamma_T T \frac{z}{z+1} 
\right), 
\eeq
\beq\label{c}
c(z)
=
\sqrt{(1-\tau)\tau} 
%\cdot
\frac{2 \tau_s(z) \gamma_T \sqrt{T}}{\sqrt{\zeta_0}}
%\cdot
\frac{z}{(z+1)(z+2)}
%\cdot
, 
\eeq
Here 
$\tau$ is the transmittance of TBS in Fig. \ref{fig:Exp_setup}, 
$\tau_h$ is the effective transmittance of the homodyne channel, 
and 
$\tau_s(z)=\tau_{s0}-\kappa z^2$ 
is a phenomenological model for a squeezing-level-dependent loss, 
which is necessary to explain the degradation of the negativity 
of $W(0,0)$ as $z$ increases.

%%%%%%%%%%%%%%%%%%%%%%%%%%%%%%%%%%%%%%%%%%%%%%%%%%%%%%%%%%%%%%%

\end{document}